\documentclass[onecolumn,a4paper,fleqn,usenatbib]{mnras}

\usepackage{newtxtext,newtxmath}
\usepackage[T1]{fontenc}
\usepackage{ae,aecompl}

\usepackage{graphicx}	% Including figure files
\usepackage{amsmath}	% Advanced maths commands

\newcommand{\ord}[1]{\textrm{ord}\left(#1\right)}
\newcommand{\abs}[1]{\left\vert#1\right\vert}

\title[Asymptotics of the Zeldovich Power Spectrum]{On the asymptotic behaviour of cosmic density-fluctuation power spectra of cold dark matter}

\author[S. Konrad et al.]{
Sara Konrad,$^{1}$\thanks{fz002@uni-heidelberg.de}
Yonadav Barry Ginat,$^{2}$\thanks{ginat@campus.technion.ac.il}
and Matthias Bartelmann$^1$
\\
$^1$ Institute for Theoretical Physics, Heidelberg University, Germany\\
$^2$ Faculty of Physics, Technion -- Israel Institute of Technology, Haifa, 3200003, Israel \\
}

\date{Accepted XXX. Received YYY; in original form ZZZ}

\pubyear{2022}

\begin{document}
\label{firstpage}
\pagerange{\pageref{firstpage}--\pageref{lastpage}}
\maketitle

% Abstract of the paper
\begin{abstract}
We study the small-scale asymptotic behaviour of the cold dark matter density fluctuation power spectrum in the Zel'dovich approximation, without introducing an ultraviolet cut-off. Assuming an initially correlated Gaussian random field and spectral index $0 < n_s < 1$, we derive the small-scale asymptotic behaviour of the initial momentum-momentum correlations. This result is then used to derive the asymptotics of the power spectrum in the Zel'dovich approximation. Our main result is an asymptotic series, dominated by a $k^{-3}$ tail at large wave-numbers, containing higher-order terms that differ by integer powers of $k^{n_s-1}$ and logarithms of $k$.
Furthermore, we show that dark matter power spectra with an ultraviolet cut-off develop an intermediate range of scales where the power spectrum is accurately described by the asymptotics of dark matter without a cut-off. These results reveal information about the mathematical structure that underlies the perturbative terms in kinetic field theory and thus the non-linear power spectrum. We also discuss the sensitivity of the small-scale asymptotics to the spectral index $n_s$.
\end{abstract}

% Select between one and six entries from the list of approved keywords.
% Don't make up new ones.

%%%%%%%%%%%%%%%%%%%%%%%%%%%%%%%%%%%%%%%%%%%%%%%%%%

%%%%%%%%%%%%%%%%% BODY OF PAPER %%%%%%%%%%%%%%%%%%

\section{Introduction}
\label{sec:1}
The Zel'dovich approximation \citep{Zeldovich1970} is ubiquitous in the theory of structure formation (see \citealt{White2014} for a relatively recent review), as a key analytic approximation that encapsulates many of the physical effects pertinent to structure formation. One of its many advantages is that it allows for the derivation of a simple formula for the two-point correlation function \citep{SchneiderBartelamnn1995,TaylorHamilton1996}, and thus the power spectrum.
While the Zel'dovich power spectrum accurately describes structures on linear scales, it is a non-linear power spectrum since it is a non-linear functional of the initial power spectrum.

In kinetic field theory (hereafter KFT), the Zel'dovich power spectrum plays an important role and is key to the full non-linear power spectrum
\citep{Bartelmannetal2017,Bartelmannetal2019, Bartelmannetal2021, Konradetal2022}. It represents the leading order power spectrum of perturbation theory and additionally is the starting point of a mean-field theory within KFT, which is an alternative way to incorporate those particle interactions that are not already captured by the Zel'dovich approximation \citep{Bartelmannetal2021}.
To derive the Zel'dovich power spectrum from KFT, three ingredients are necessary: (i) the specification of the initial conditions in classical phase-space, (ii) the assumption of straight particle trajectories\footnote{They are inertial if time is transformed to a specific non-linear function of itself.} that contain parts of the gravitational interactions and (iii) the expanding space-time as background \citep{Bartelmannetal2016,Bartelmannetal2017}.

We consider Gaussian initial conditions, implying that two-point correlation functions are sufficient to completely determine the initial probability distribution in phase space.
We assume that initial momenta are correlated, while initial density-density and density-momentum correlations are neglected. This is reasonable on large scales and at late times, as they grow much more slowly than the initial momentum-momentum correlations.
The initial velocity field can be considered to be curl-free, such that it can be expressed in terms of a gradient of a velocity potential $\psi$. Since the continuity equation has to hold initially, the initial density contrast has to satisfy Poisson's equation
\begin{equation}
\delta^\mathrm{(i)} = - \nabla^2 \psi\;.
\end{equation}
(All the constants that might appear are absorbed into $\psi$ here.)
Consequently, the initial velocity potential power spectrum is related to the initial density fluctuation power spectrum via
\begin{equation}
  P_{\psi}^\mathrm{(i)}(k) = k^{-4}\,P_\delta^\mathrm{(i)}(k)\;.
\label{eq:5}
\end{equation}
The initial momentum-momentum correlation matrix of two particles separated by $\mathbf{q}$ is then given by
\begin{align}
  \hat{C}_{pp}\left(\mathbf{q}\right) = \int_{k'}\left(
    \mathbf{k}'\otimes \mathbf{k}'\,
  \right)P_{\psi}^\mathrm{(i)}(k')\,\mathrm{e}^{\mathrm{i} \mathbf{k}'\cdot \mathbf{q}}\;.
\label{eq:2}
\end{align}
The ansatz that the particle trajectories are straight is key to the Zel'dovich approximation.
In KFT, this ansatz is expressed by splitting up the actual time dependent trajectory of a particle into the Zel'dovich part $\mathbf{q}^\mathrm{Z}(t)$, that is linear in the initial phase space coordinates $\mathbf{q}^{(i)}$, $\mathbf{p}^\mathrm{(i)}$, and an interaction part $\mathbf{q}^\mathrm{I}(t)$ that contains all the remaining interactions  that are not captured by the Zel'dovich approximation,
\begin{equation}
\mathbf{q}(t)
= \mathbf{q}^{(i)} + g^\mathrm{Z}_{qp}(t,t^{(i)}) \mathbf{p}^{(i)} + \mathbf{q}^{I}(t)
\equiv \mathbf{q}^\mathrm{Z}(t) + \mathbf{q}^\mathrm{I}(t)\;,
\label{eq:Zeldovich trajectories}
\end{equation}
where $g_{qp}(t,t^{(i)})$ is the so-called propagator. Considering only $\mathbf{q}^{Z}(t)$ and identifying $g^\mathrm{Z}_{qp}(t,t^{(i)}) \equiv D_+(t)-D_+(t^{(i)})$ with the linear growth factor $D_+$, yields the Zel'dovich approximation. In the following, we define the initial time $t^{(i)} = 0$ and then replace in our notation the propagator $g_{qp}(t,0)$ by $t$. This not only simplifies the notation but also highlights that the propagator serves as a time coordinate. We emphasise that Eq. \eqref{eq:Zeldovich trajectories} is exact.

When these assumptions are combined, the unperturbed zeroth-order power spectrum of KFT is equivalent to the Zel'dovich power spectrum and is given by
\begin{equation}
  \mathcal{P}(k,t) = \mathrm{e}^{-Q}\int\mathrm{d}^3q\left(
    \mathrm{e}^{t^2\,\mathbf{k}^\top\hat{C}_{pp}(\mathbf{q})\mathbf{k}}-1
  \right)\mathrm{e}^{\mathrm{i} \mathbf{k}\cdot \mathbf{q}}\;,
\label{eq:KFT zeldovich PS}
\end{equation}
as shown in \cite{Bartelmannetal2016,Bartelmannetal2017}.
The exponent $Q$ is
\begin{equation}
  Q = \frac{\sigma_1^2}{3}\,k^2\,t^2\;,
\label{eq:3}
\end{equation}
where $\sigma_1^2$ is one of the moments
\begin{equation}
  \sigma_n^2 = \frac{1}{2\pi^2}\int_0^\infty\mathrm{d} k\,k^{2n+2}\,
  P_{\psi}^{(i)}(k)
\label{eq:4}
\end{equation}
of the power spectrum $P_{\psi}^{(i)}$ of the initial velocity potential $\psi$. It is important for the present derivation of our results that $\sigma_2$ is finite, which is the case for $0<n_s<1$, but higher moments are allowed to be infinite.

We emphasise that nothing in this paper relies on features peculiar to KFT -- indeed, all that we assume is that the power spectrum in the Zel'dovich approximation is given by \eqref{eq:KFT zeldovich PS}, which can be reached from, e.g., Lagrangian perturbation theory, too \citep{TaylorHamilton1996}.

The small-scale limit of the power spectrum is of interest to cosmologists for many reasons: it can be used to study the behaviour of dark matter inside haloes \citep{MaFry2000}, and thereby to gain some understanding on their universal density profile (see, e.g., \citealt{Wangetal2020}). Besides, the rate of decay of the power spectrum with $k$ at large $k$ influences the degree of relativistic gravitational back-reaction (see, e.g., \citealt{Adameketal2018,Ginat2021}). Another reason for investigating the power spectrum on small scales is that, through the time dependence of the asymptotic expansion, it might shed light on the process of virialisation in gravitational collapse \citep{Peebles1980}. %\sk{To Matthias: Would you write here or somewhere suitable a sentence on why the Zel'dovich power spectrum is interesting from a purely statistical physics point of view?}
Even though we are deriving results here for ensembles of classical particles following Zel'dovich trajectories in an expanding background space-time, we should like to emphasise that our results are applicable without change also to static space-time and inertial trajectories in a different time coordinate. Our essential assumptions are that initial phase-space coordinates sample a Gaussian random field whose power spectrum satisfies some fairly general properties. Our results should therefore extend to wide classes of classical particles streaming freely from Gaussian initial conditions.

\cite{Konradetal2020} proved that \eqref{eq:KFT zeldovich PS} may be expanded as
\begin{equation}
\mathcal{P}(k,t) \sim \sum_{m=0}^{\infty} \frac{\mathcal{P}^{(m)}(t)}{k^{3+2m}}\;,
\label{eq:6}
\end{equation}
for $k \rightarrow \infty$, where the time dependent coefficients $\mathcal{P}^{(m)}$ depend on the initial power spectrum only through $\sigma_n^2$ for $n\leq m$, and they derived explicit expressions for these coefficients. \cite{ChenPietroni2020} also derived a similar expansion, based on a Lagrangian view-point. However, these results relied on the assumption that \emph{all} the moments \eqref{eq:4} are finite, which implies that if one assumes dispersion free cold dark matter, where the initial power spectrum has a flat tail, one needs to introduce a regulator which then affects all results. In this work, we derive the asymptotic behaviour for cold dark matter power spectra, without introducing a regulator, which leads to markedly different results.

We do so by applying the Mellin transform technique of asymptotic expansions, which consists of computing the Mellin transform, continuing it analytically to the entirety of the complex plane (except for various possible poles), and then using this analytic continuation, together with Gauss' residue theorem to compute the inverse Mellin transform by shifting the contour of integration (see, e.g. Chapter 4 of \citealt{bleistein1975asymptotic} or Appendix A of \citealt{BarryHughes1995}). This technique has been applied before in various works in astrophysical contexts (e.g., \citealt{ToumaTremaine1997,Ginatetal2019}, to which we refer the readers for pedagogical utilisations). Its great advantage over other expansion techniques is that it produces, automatically, the correct expansion functions, even non-analytic ones: it can handle non-integer powers in a power-series, logarithms, \emph{etc}. For completeness, we summarise the relevant theorem in Appendix \ref{appendix:Mellin}.

This paper is structured as follows: we start by deriving the asymptotic expansions of the initial momentum correlation matrix $\hat{C}_{pp}$ both at small and large $q$ in \S \ref{sec:initial momentum correlations}. We then use these results in conjunction with the Mellin transform technique in \S \ref{sec:ps large scale} to expand \eqref{eq:KFT zeldovich PS} at large values of $k$, without assuming an ultraviolet cut-off on the initial power spectrum. In \S \ref{sec:late time} we specialise the expansion to late times (we define what `late' means there), where some simplifications are possible. We then study the time-dependence of the expansion coefficients in \S \ref{sec:time evolution of coefficients}, and summarise in \S \ref{sec:summary}.

\section{Small-scale asymptotics of the initial momentum correlations}
\label{sec:initial momentum correlations}
We start with the small-scale asymptotics of the initial momentum-momentum correlations of two particles that are separated by a vector $\mathbf{q}$. As shown in \cite{Bartelmannetal2017}, the initial correlation matrix can be written as
\begin{equation}
\hat{C}_{pp}(\mathbf{q}) = -  \hat{\mathbf{q}} \otimes \hat{\mathbf{q}} \; a_2(q) - \mathbb{I}_3 a_1(q)\,,
\label{eq:7}
\end{equation}
with $\hat{\mathbf{q}}$ being the unit vector pointing in the direction of $\mathbf{q}$ and
\begin{align}
a_1(q) & \equiv \frac{\xi'_{\psi}(q)}{q} = - \frac{1}{2\pi^2} \int_0^{\infty} \mathrm{d}k \ P^{\text{(i)}}_{\delta} (k) \frac{j_1(kq)}{kq}\;,
\label{eq:a1a2:11}
\\
a_2(q) & \equiv \xi''_{\psi}(q) -  \frac{\xi'_{\psi}(q)}{q} = \frac{1}{2\pi^2} \int_0^{\infty} \mathrm{d}k \ P^{\text{(i)}}_{\delta} (k) j_2(kq)\;,
\label{eq:a1a2:12}
\end{align}
where $\xi_{\psi}$ denotes the correlation function of the initial velocity potential and $j_n$ denotes the spherical Bessel function of order $n$.
We take the initial power spectrum to be
\begin{equation}
    P_\delta^{(i)}(k)
    = \frac{8 \pi^2 A_s \left(\frac{c}{H_0} \right)^4 k_s^{n_s}}{25 \Omega_m^2 k_0^{n_s-1}} \left[\frac{k}{k_s}\right]^{n_s} \tilde T_D^2\left(\frac{k}{k_{\rm eq}}\right)
    \equiv A \left[\frac{k}{k_s}\right]^{n_s} \tilde T_D^2\left(\frac{k}{k_{\rm eq}}\right)\;,
\end{equation}
where the wave number $k_s$ is defined below, and use a Dicus transfer function \citep[p. 307]{Weinberg2008}
\begin{equation}
    \tilde T_D^2(\kappa) = \left(\frac{\ln \left((\tilde c_1 \kappa )^2+1\right)}{(\tilde c_1 \kappa )^2}\right)^2 \frac{(\tilde c_2 \kappa )^2+(\tilde c_3 \kappa )^4+(\tilde c_4 \kappa )^6+1}{(\tilde c_5 \kappa )^2+(\tilde c_6 \kappa )^4+(\tilde c_7\kappa )^6+1}\;,
\label{eq:Dicus}
\end{equation}
with $\tilde c_1 = 0.124$, $\tilde c_2 = 1.257$, $\tilde c_3 = 0.4452$, $\tilde c_4 = 0.2197$, $\tilde c_5 = 1.606$, $\tilde c_6 = 0.8568$, and $\tilde c_7 = 0.3927$.

We use the Planck 2016 cosmological parameters \citep{Planck2015}, as given in Tab.~\ref{tab:parameters}. We stress that all results presented here are independent of the explicit choice of a transfer function for the initial power spectrum. The only important features are the asymptotics at small scales, i.e. at $k \to \infty$, that has to agree with the general form specified in \eqref{eq:13} below; and that $0 < n_s < 1$.
\begin{table}
    \centering
    \begin{tabular}{|c|c|c|c|c|c|c|c|c|}
    \hline
    Parameter & $h$ & $\Omega_m$ & $z_{\rm eq}$ & $\Omega_r$ & $k_0$ & $A_s$ & $k_{\rm eq}$ & $n_s$ \\
    \hline
    Value & $0.6774$ & $0.3089$ & $3371$ & $\frac{\Omega_m}{1+z_{\rm eq}}$ & $0.05~ \textrm{Mpc}^{-1}$ & $2.142 \times 10^{-9}$ & $\frac{\Omega_m H_0}{c\sqrt{\Omega_r}}$ & $0.9667$ \\
    \hline
    \end{tabular}
    \caption{Cosmological parameters used in this paper \citep{Planck2015}.}\label{tab:parameters}
\end{table}

To derive an asymptotic expansion, it is best to move from distances $q$ to a dimension-less variable $x$, defined by $x = qk_s$, where
\begin{equation}
    k_s = \frac{k_{\rm eq}}{c_1},
    \label{eq:12}
\end{equation}
and switch from wave numbers $k$ to the dimension-less quantity $\kappa = k/k_s$. This effectively re-scales $\tilde c_{1,\ldots,7} \mapsto c_{1,\ldots,7} \equiv {\tilde c_{1,\ldots,7}}/{\tilde c_1}$, which is what we shall do below. Likewise, we define $T_D(\kappa)$ in the same way as $\tilde T_D$, but with $c_{1,\ldots,7}$ instead of $\tilde c_{1,\ldots,7}$.

The initial power spectrum $P_{\delta}^{(i)}$ is taken to have an asymptotic expansion as $\kappa \equiv \frac{k}{k_s} \to \infty$ that reads
\begin{equation}
P_{\delta}^{(i)} \sim \kappa^{n_s-4} \sum_{m=0}^{\infty} \kappa^{-m} \sum_{n=0}^{2} c_{mn} \ln^n \kappa\;.
\label{eq:13}
\end{equation}
Applying this to \eqref{eq:Dicus} , one has
\begin{equation}
    c_{02} = \frac{4Ac_4^6}{c_7^6},
\end{equation}
and $c_{00} = c_{01} = 0$. We will not need the higher-order terms in this paper.

Applying the Mellin transform technique (cf. Appendix \ref{appendix:Mellin}) to the definitions of $a_{1,2}$, we find that, as $x = q k_s \to 0$,
\begin{align}
    \label{eqn:a1 at small q}a_1(q) & \sim -\frac{1}{6\pi^2} \mathcal{M}\left[P_\delta^{(i)};1\right] + \frac{x^2}{60\pi^2}\mathcal{M}\left[P_\delta^{(i)};3\right]
    -\frac{x^{3-n_s}}{2\pi^2} \frac{c_{02}}{A}\sum_{j=0}^2 \binom{2}{j}\left(-\ln x\right)^j \mathcal{M}^{(2-j)}\left[j_1;n_s-4\right] + o(x^{3-n_s})
    \\
    \label{eqn:a2 at small q} a_2(q) & \sim \frac{x^2}{30\pi^2}\mathcal{M}\left[P_\delta^{(i)};3\right]
    + \frac{x^{3-n_s}}{2\pi^2} \frac{c_{02}}{A}\sum_{n=0}^2 \binom{2}{j}\left(-\ln x\right)^j \mathcal{M}^{(2-j)}\left[j_2;n_s-3\right] + o(x^{3-n_s}),
\end{align}
where $\mathcal{M}^{(k)}[f;z]$ is the $k$-th derivative of the Mellin transform of the function $f$, $\mathcal{M}[f;z]$, with respect to $z$. Note that the functions $a_{1,2}$ have units of area. Here, we have defined
\begin{equation}
    \mathcal{M}\left[P_\delta^{(i)};z\right] \equiv Ak_s \int_0^\infty \kappa^{n_s + z-1} T_D^2(\kappa) \mathrm{d}\kappa\;,
    \label{eq:17}
\end{equation}
which also has units of area.

The coefficient $\xi_1(x,n_s)$ of $x^{3-n_s}$ in $a_1$ can be written as
\begin{equation}
    \begin{aligned}
    \xi_1(x,n_s) &= -\frac{2}{\pi^2}\frac{c_4^6}{c_7^4}\left(\mathcal{M}''[j_1;n_s-4] - 2 \ln x \mathcal{M}'[j_1;n_s-4] + \ln^2x \mathcal{M}[j_1;n_s-4]\right)
    \\
    & =:
    \xi_{10}(n_s) + \ln x \ \xi_{11}(n_s) + \ln^2 x \ \xi_{12}(n_s)\;,
    \end{aligned}
    \label{equation::20}
\end{equation}
and the coefficient $\xi_2(x,n_s)$ of $x^{3-n_s}$ in $a_2$ is likewise
\begin{equation}
    \begin{aligned}
    \xi_2(x,n_s) &= \frac{2}{\pi^2}\frac{c_4^6}{c_7^4}\left(\mathcal{M}''[j_2;n_s-3] - 2 \ln x \mathcal{M}'[j_2;n_s-3] + \ln^2x \mathcal{M}[j_2;n_s-3]\right)
    \\
    &=:
    \xi_{20}(n_s) + \ln x \ \xi_{21}(n_s) + \ln^2 x \ \xi_{22}(n_s)\;
    \;.
    \end{aligned}
    \label{equation::21}
\end{equation}

\begin{figure}
    \includegraphics[width=1.0\textwidth]{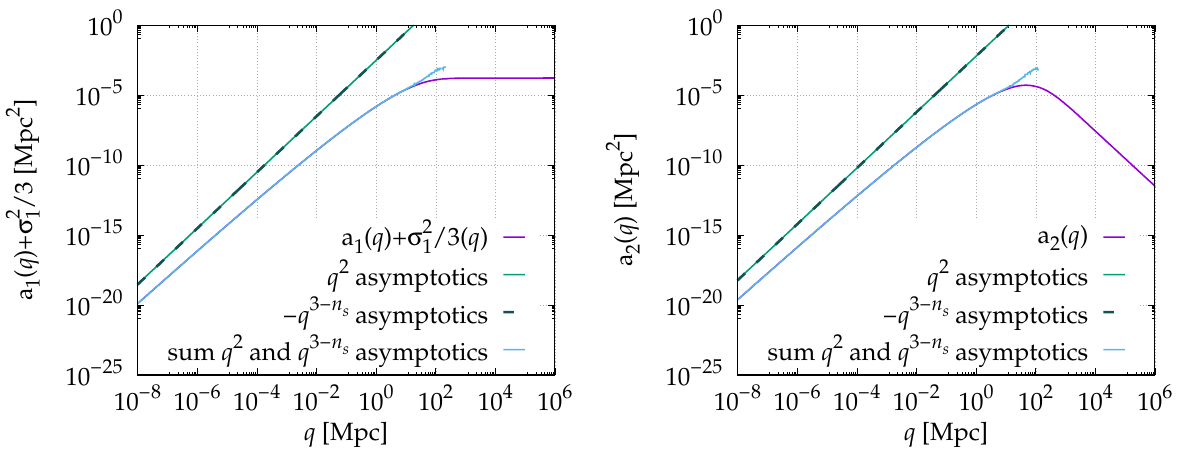}
    \caption{Left: The function $a_1(q)+\frac{\sigma_1^2}{3}$ (purple line) is shown with $a_1$ from \eqref{eq:a1a2:11} together with the order $q^2$ asymptotics (green line) and the absolute value of the order $q^{3-n_s}$ asymptotics including the both log terms (blue dashed line) from \eqref{eqn:a1 at small q} as well as the sum of those first two leading order asymptotics (light blue line). Right: Same as left, but $a_2(q)$ from \eqref{eq:a1a2:11} with asymptotics from \eqref{eqn:a2 at small q}. The Dicus transfer function \citep{Weinberg2008} and the parameters in Table \ref{tab:parameters} are used.}
    \label{fig:a1a2}
\end{figure}

\begin{figure}
    \includegraphics[width=1.0\textwidth]{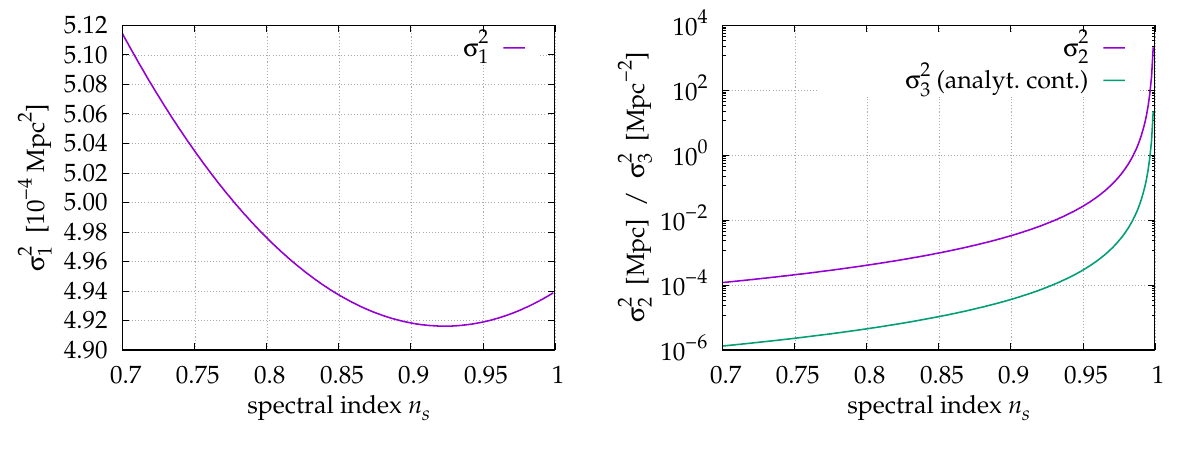}
    \caption{The moments of the initial velocity potential as a function of the spectral index $n_s$ for Planck parameters (Table \ref{tab:parameters}). While the first moment $\sigma_1^2$ (left) varies only little with the spectral index, the second moment $\sigma_2^2$ and the analytic continuation of the third moment $\sigma_3^2$ diverge as $n_s \rightarrow 1$ (right). The Dicus transfer function \citep{Weinberg2008} is used.}
    \label{fig:sigmas_for_ns}
\end{figure}

Figure \ref{fig:a1a2} shows that considering only the order $q^2$ asymptotics (green lines) is insufficient to describe the functions $a_{1,2}$ (purple lines) accurately  at cosmologically reasonable small scales. Since the next order, the $\ord{q^{3-n_s}}$ terms (dashed dark blue lines), are negative and $\abs{1-n_s} \ll 1$, they balance the order $q^2$ terms. The sum of both orders (light blue lines) describes the small scale asymptotics of $a_{1,2}$ very well up to Mpc scales. Let us recall here, that our results \eqref{eqn:a1 at small q} and \eqref{eqn:a2 at small q} are only valid for $n_s < 1$. In the  limit $n_s \rightarrow 1$, the analytical continuation of $\mathcal{M}\left[P_{\delta}^{(i)};3\right]$, i.e. the second moment of the initial velocity potential power spectrum, diverges (see Fig. \ref{fig:sigmas_for_ns}). Thus, for $n_s=1$, the resulting series contain terms with even powers of $q$ and logarithms of higher order than 2 due to coinciding poles of $\mathcal{M}\left[P_{\delta}^{(i)};z\right]$ and $\mathcal{M}\left[j_{1,2};z\right]$. The coefficients may be computed by applying the second strategy of Case IV of the Mellin transform technique described in Appendix \ref{appendix:Mellin}.

\section{Small-scale asymptotics of the Zel'dovich power spectrum}
\label{sec:ps large scale}
The small-scale asymptotics of the Zel'dovich power spectrum as $k\to \infty$ are determined by the saddle points of the exponent of the integrand in \eqref{eq:KFT zeldovich PS}.
The saddle point, that dominates the $k \to \infty$ asymptotics of the integral, is located at the origin $q \to 0^+$.

To prove this claim, let us now show that any potential saddle point of the form $q = \ord{k^{\beta}}$, where $\beta > 0$, gives an exponentially-decreasing contribution to $\mathcal{P}(k,t)$.
As $q \to \infty$, the Mellin-transform technique may be used to show that $a_{1,2} \sim q^{-1-n_s}$. (The explicit expression for the large-scale asymptotics of $a_{1,2}$ is shown in Appendix \ref{appendix:asymptotics large scale}.) This result implies that for any saddle point where $\displaystyle \lim_{k \to \infty} q = \infty$, we have $\hat{C}_{pp}(q) = o(1)$, and therefore that the exponential,
\begin{equation}
    t^2k^i(\hat{C}_{pp})_{ij}k^j - Q\;,
\end{equation}
with $Q \sim k^2$ from \eqref{eq:3} is dominated by $-Q$ in the large-$k$ limit. That is, the contribution to $\mathcal{P}$ from such a saddle point is proportional to the exponentially-decaying factor
\begin{equation}
    \exp\left[ -k^2\frac{\sigma_1^2t^2}{3} + o\left(k^2\right)\right].
\end{equation}
In a similar fashion, when $\beta = 0$ and $q \neq 0$ (i.e. $q$ has a constant finite value), $\frac{t^2k^i(\hat{C}_{pp})_{ij}k^j - Q}{k^2}$ is a negative, order unity quantity, and thus gives rise to an exponentially-decaying contribution. Consequently, if a saddle point exists with $\beta < 0$, i.e. at $q \to 0^+$, and if this saddle point yields a contribution that decays more slowly than exponentially, then it does dominate over all other saddle points with $\beta \geq 0$.

We are now in a position to evaluate the dominant contribution to the power spectrum, by inserting \eqref{eqn:a1 at small q} and \eqref{eqn:a2 at small q} into \eqref{eq:7} and \eqref{eq:KFT zeldovich PS}, and integrating over $q$. Explicitly, upon defining $\sigma^2 \equiv \mathcal{M}[P_\delta^{(i)};3]$ (having dimensions of area; see \eqref{eq:17}), we obtain, as $k \rightarrow \infty$,
\begin{equation}
    \begin{aligned}
    \mathcal{P}(k,t) & \sim \int \mathrm{d}^3q~\mathrm{e}^{-t^2(\mathbf{k}\cdot\hat{\mathbf{q}})^2\left(x^2 \frac{\sigma^2}{15} + x^{3-n_s} \xi_2\right) - t^2k^2\left(x^2 \frac{\sigma^2}{30}  - x^{3-n_s} \xi_1\right) + \mathrm{i}\mathbf{k}\cdot\mathbf{q}
    } \\ &
    \sim 2\pi k_s^{-3} \int_{-1}^1 \mathrm{d}\mu \int_0^{\infty} \mathrm{d}x~ x^2 e^{-t^2 k^2 (2\mu^2 + 1) \frac{\sigma^2}{30} x^2 + \mathrm{i}\mathbf{k}\cdot\mathbf{q}} \exp\left(t^2k^2 x^{3-n_s}(\xi_1 - \mu^2\xi_2)\right)\;,
    \end{aligned}
\end{equation}
where we changed variables to the dimension-less radial coordinate $x=k_s q$. Recall that the functions $\xi_{1,2}$ both are quadratic polynomials in logarithms of $x$. Defining $\xi \equiv \xi_1-\mu^2 \xi_2$ and using the Taylor expansion of the last exponential to generate an asymptotic series expression, we arrive at
\begin{equation}
    \begin{aligned}
    \mathcal{P}(k,t) &
    \sim 2\pi k_s^{-3}\sum_{m=0}^\infty \int_{-1}^1 \mathrm{d}\mu \int_0^{\infty} \mathrm{d}x~ x^2 e^{-t^2 k^2 (2\mu^2 + 1)\frac{\sigma^2}{30} x^2}\cos\left(kq\mu\right) \frac{(t k)^{2m} x^{m(3-n_s)}}{m!}\xi^m \\ &
    = \frac{2\pi}{(tkk_s\sigma)^3}\sum_{m=0}^\infty \int_{-1}^1 \mathrm{d}\mu \int_0^{\infty} \mathrm{d}y~ y^2 e^{-(2\mu^2 + 1)y^2/30}\cos\left(\frac{y\mu}{t\sigma k_s}\right) \frac{ y^{m(3-n_s)}}{m!(t k \sigma)^{m(1-n_s)}}\left[\frac{\xi}{\sigma^2}\right]^m\;,
    \label{eq:23}
    \end{aligned}
\end{equation}
as $k \to \infty$. Again, recall that $\frac{\xi}{\sigma^2}$ is a dimension-less quantity. The (dimension-less) integration variable $y$ is defined as $y=tx\sigma k$.
So, as $k \to \infty$,
\begin{equation}
    \mathcal{P}(k,t) \sim \frac{2\pi}{(tkk_s\sigma)^3}\sum_{m=0}^\infty \int_{-1}^1 \mathrm{d}\mu \int_0^{\infty} \mathrm{d}y~ y^2 \cos\left(\frac{y\mu}{t\sigma k_s}\right) \frac{e^{-(2\mu^2 + 1)y^2/30} y^{m(3-n_s)}}{m!(t k \sigma)^{m(1-n_s)}}\left[\frac{\xi}{\sigma^2}\right]^m + O\left(k^{-5}\right).
\end{equation}
One can summarise this expansion as
\begin{equation}\label{eqn:P_z m series}
    \mathcal{P}(k,t) = \frac{2\pi}{(k \tau_2)^3}\sum_{m=0}^\infty \frac{k_s^{m(1-n_s)}\mathcal{P}_m\left(t,\ln(k\tau_2/k_s)\right)}{(k\tau_2)^{m(1-n_s)}},
\end{equation}
with $\tau_2 \equiv t\sigma k_s$ and\footnote{Recall that $\tau_2$ is simply a monotonically increasing function of time.}
\begin{equation}
\label{eqn:P_m definition}
    \mathcal{P}_m\left(t,\ln(k\tau_2/k_s)\right) = \int_{-1}^1 \mathrm{d}\mu \int_0^{\infty} \mathrm{d}y~ y^2 \cos\left(\frac{y\mu}{\tau_2}\right) \frac{e^{-(2\mu^2 + 1)y^2/30 } y^{m(3-n_s)}}{m!}\left[\frac{\xi}{\sigma^2}\right]^m.
\end{equation}
We are now in a position to extract all of the $k$-dependence of the power spectrum. Like $\xi_{1,2}$, $\xi$ is a quadratic polynomial in $\ln k$, and therefore
\begin{equation}\label{eqn:whole k dependence extracted}
\mathcal{P}(k,t) \sim \frac{1}{(k \tau_2)^3}\sum_{m=0}^{\infty} \left(\frac{k_s}{k \tau_2} \right)^{m(1-n_s)} \sum_{n=0}^{2m} \mathcal{P}_{mn}(t) \ln^{n} \left(\frac{k \tau_2}{k_s}\right) + \mathcal{O}\left(k^{-5}\right)
\end{equation}
with
\begin{equation}
\mathcal{P}_{mn}(t) = \frac{2 \pi}{m! (\sigma_2^2)^m}  \int_{-1}^1 \mathrm{d}\mu \int_0^{\infty} \mathrm{d}y \;
\mathrm{e}^{-\frac{1}{30} (2 \mu^2 +1)y^2 } \mathrm{e}^{\mathrm{i}\mu \frac{y}{\tau_2}}
y^{m(3-n_s)+2} \xi_{m,n}\left(\mu^2, \ln y \right)\;,
\label{eq:28}
\end{equation}
with $\xi_{m,0}\left(\mu^2, \ln y \right) \equiv \tilde{\xi}^m_0\left(\mu^2, \ln y \right)$ and for $n \neq 0$, the functions $\xi_{m,n}$ are given recursively by
\begin{equation}
\xi_{m,n}\left(\mu^2, \ln y \right) = \frac{1}{n \tilde{\xi}_0} \left[(m-n+1) \tilde{\xi}_1 \xi_{m,n-1} + (2m - n + 2) \tilde{\xi}_2 \xi_{m,n-2} \right]\;,
\end{equation}
with
\begin{align}
\tilde{\xi}_2(\mu^2, \ln y)
& \equiv
\xi_{12} - \mu^2 \xi_{22}\;,
\\
\tilde{\xi}_1(\mu^2, \ln y)
& \equiv
- \left( \xi_{11} - \mu^2 \xi_{21} \right) - 2 \ln y \ \left( \xi_{12} - \mu^2 \xi_{22} \right)\;,
\\
\tilde{\xi}_0(\mu^2, \ln y)
& \equiv
\ln^2 y \ \left(  \xi_{12} - \mu^2 \xi_{22} \right)
+ \ln y \ \left(  \xi_{11} - \mu^2 \xi_{21} \right)
+  \xi_{10} - \mu^2 \xi_{20}\;,
\end{align}
with $\xi_{10}$, $\xi_{11}$, $\xi_{12}$, $\xi_{20}$, $\xi_{21}$, and $\xi_{22}$ given by Eqs.~\eqref{equation::20} and \eqref{equation::21}.

Equation \eqref{eqn:whole k dependence extracted} is the main result of this paper: it shows the full dependence of the Zel'dovich power spectrum on $k$, for large values of $\frac{k}{k_s}$, up to corrections of order $k^{-5}$ (for the Dicus transfer function), which are in principle calculable using the Mellin transform technique. Here, as opposed to earlier works \citep{Konradetal2020, ChenPietroni2020}, higher moments of the initial power spectrum need not be finite. 
There, a UV cutoff of the initial power spectrum had to be introduced. In its presence, a full asymptotic expansion for $k\rightarrow \infty$ is given by \eqref{eq:6}. Without a UV cutoff in the initial power spectrum, the asymptotics of the evolved Zel'dovich power spectrum is instead given by \eqref{eqn:whole k dependence extracted} up to $\mathcal{O}\left(k^{-5}\right)$. With a UV cutoff, \eqref{eqn:whole k dependence extracted} reduces to the single non-vanishing term with $m = 0$, expressing the leading-order $k^{-3}$ asymptotics of the Zel'dovich power spectrum. We emphasize that terms with $m>0$ are not only negligible, but strictly absent in presence of a UV cutoff since they only appear due to the poles of the Mellin transform of the initial power spectrum. Our result \eqref{eqn:whole k dependence extracted} also shows why terms beyond the order $k^{-3}$ are necessary to describe the Zel'dovich power spectrum on cosmologically relevant scales: the weak increase of logarithms in the numerator almost precisely cancels the weak increase of $k^{m(1-n_s)}$ (for realistic $n_s$) in the denominator. The relevance of high orders can be seen in Fig. \ref{fig:late time z=0}, which shows the asymptotic series up to a maximum value of $m \leq M$, for various $M$. This maximum value $M$ is chosen such that contributions way beyond order $k^{-5}$ are considered, indicating that the missing $\mathcal{O}(k^{5})$-terms in \eqref{eqn:whole k dependence extracted} are less relevant; see also Appendix \ref{appendix: neglect k^-5}. We will discuss this in more detail for redshift $z=0$ in the next section.

\begin{figure}
    \includegraphics[width=1.0\textwidth]{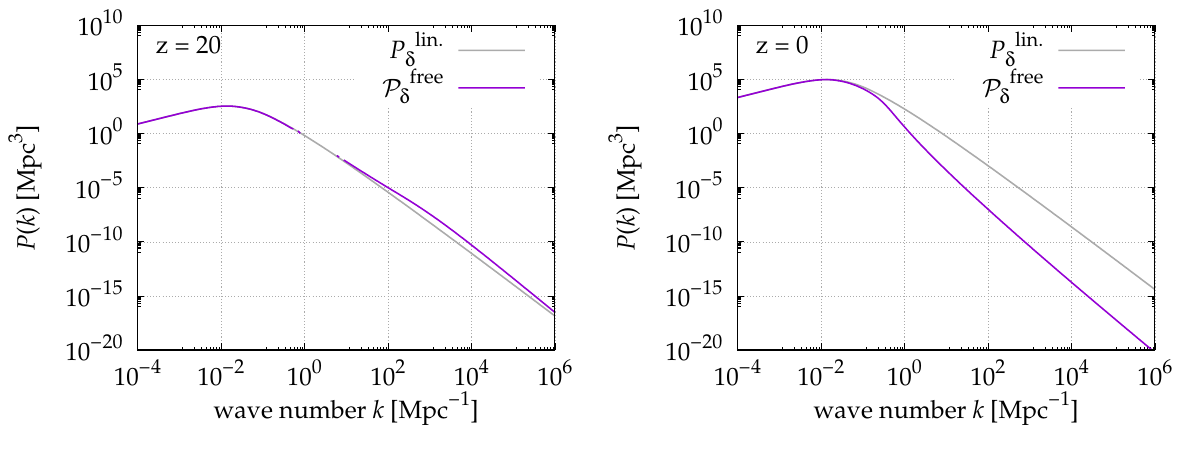}
    \caption{Left: The Zel'dovich power spectrum (purple line) from \eqref{eq:KFT zeldovich PS} and the power spectrum from linear theory (gray line) $P_{\delta}^{\text{lin.}} = D_+^2(z) P_{\delta}^{ini}$ are shown at redshift $z = 20$. Right: Same power spectra at today's redshift $z=0$. Parameters as shown in Table \ref{tab:parameters}.}
    \label{fig:free_spectra}
\end{figure}

\section{Late time evolution}
\label{sec:late time}
\begin{figure}
    \centering
    \includegraphics[width=1.0\textwidth]{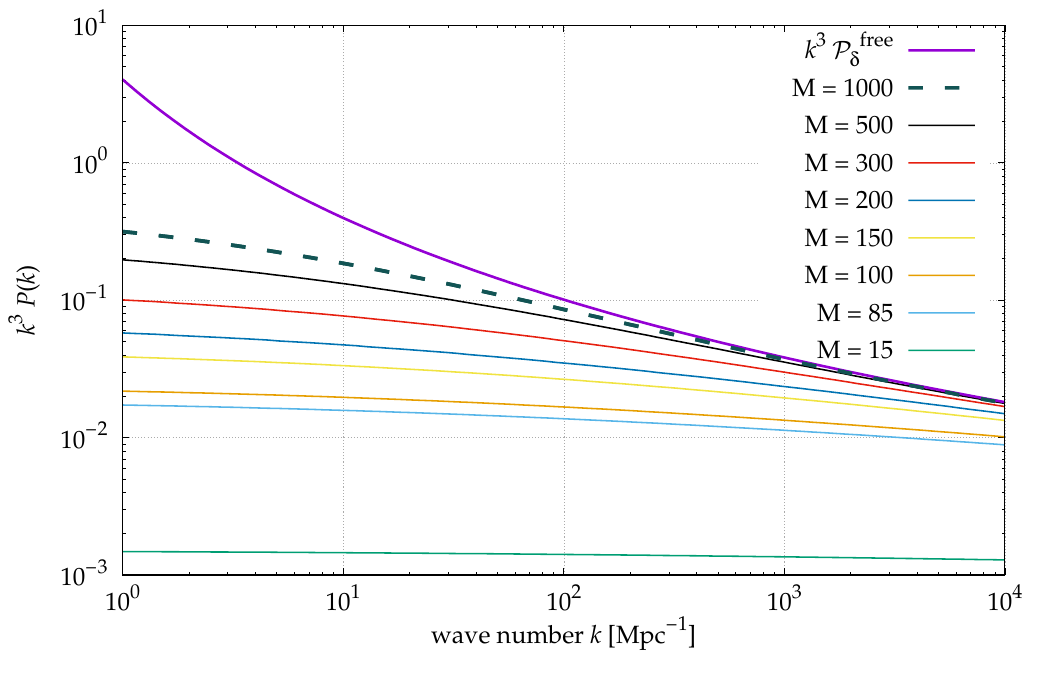}
    \caption{The dimension-less cold dark matter power spectrum in the Zel'dovich approximation (purple line) at today's redshift $z=0$, compared with the analytical prediction of \eqref{eqn:late time Pm}, up to high orders $M$. The Dicus transfer function \citep{Weinberg2008} and the parameters in table \ref{tab:parameters} are used.}
    \label{fig:late time z=0}
\end{figure}

As $t$ grows, $\tau_2$ becomes quite large. The argument of the cosine in \eqref{eqn:P_m definition} tends to $0$ at late times, thus
\begin{equation}\label{eqn:late time Pm}
    \mathcal{P}_m \to \int_{-1}^1 \mathrm{d}\mu \int_0^{\infty} \mathrm{d}y~ y^2 \frac{e^{-(2\mu^2 + 1)y^2/30}y^{m(3-n_s)}}{m!}\left[\frac{\xi}{\sigma^2}\right]^m,
\end{equation}
which is time-independent and simpler to evaluate numerically. More precisely,
in the late-time limit $\tau_2 \rightarrow \infty$, $\mathcal{P}_{mn}$ from \eqref{eq:28} can be expanded into the series
\begin{equation}
\mathcal{P}_{mn}(t) \to \frac{2 \pi}{m! (\sigma_2^2)^m}  \int_{-1}^1 \mathrm{d}\mu \int_0^{\infty} \mathrm{d}y \;
\mathrm{e}^{-\frac{1}{30} (2 \mu^2 +1)y^2 } \sum_{l=0}^{\infty} \frac{(-1)^l}{(2l)!} \left(\mu \frac{y}{\tau_2} \right)^{2l}
y^{m(3-n_s)+2} \xi_{m,n}\left(\mu^2, \ln y \right)\;.
\end{equation}
For the parameters in Table \ref{tab:parameters}, $\tau_2 \approx 259$ at redshift $z = 0$, so one can safely use the late-time expression. The results are shown in Fig. \ref{fig:late time z=0}. One can see excellent agreement with the asymptotic expansion at scales $k \ge 10^3$ Mpc$^{-1}$ when going to high orders in the asymptotics: for low numbers of terms ($M=15$, green line), the asymptotics underestimate the actual Zel'dovich power spectrum by more than an order of magnitude. Summing more and more orders improves the result significantly. Including terms up to order $M=1000$ leads to an agreement of the asymptotics with the Zel'dovich power spectrum at small scales.
We note that the computation of the double integral in \eqref{eqn:late time Pm} is more involved for large $m$, and we explain the method we used in Appendix \ref{appendix:large m}. Furthermore, let us note that the order $k^{-5}$ term can be neglected although we go to very high orders in $M$, which we discuss in Appendix \ref{appendix: neglect k^-5}.

%with
%\begin{equation}
%\mathcal{P}_{mnl}(t) = \frac{2 \pi}{m! (\sigma_2^2)^m}  \frac{(-1)^l}{(2l)! \tau_2^{2l}} \int_{-1}^1 \mathrm{d}\mu \int_0^{\infty} \mathrm{d}y \;
%\mathrm{e}^{-\frac{1}{30} (2 \mu^2 +1)y^2 } \mu^{2l}
%y^{m(3-n_s)+2(l+1)} \xi_{m,n}\left(\mu^2, \ln y \right)\;,
%\end{equation}
%\begin{equation}
%\mathcal{P}_{mn0} = \frac{2 \pi}{m! (\sigma_2^2)^m} \int_{-1}^1 \mathrm{d}\mu \int_0^{\infty} \mathrm{d}y \;
%\mathrm{e}^{-\frac{1}{30} (2 \mu^2 +1)y^2 }
%y^{m(3-n_s)+2} \xi_{m,n}\left(\mu^2, \ln y \right)\;,
%\end{equation}

\section{Implications for dark matter}
\label{sec:time evolution of coefficients}
If dark matter is made of massive fermions, then the initial power spectrum does not decay like $k^{n_s-4} \log^2 k$. It may rather have a cutoff, determined by the particle mass and the particle production mechanism, which is related to the type(s) of interaction the particle undergoes \citep{Greenetal2004}. The ultraviolet cut-off $k_D$ is also typically related to a temperature of dark matter.
In this section, we discuss two aspects of the Zel'dovich power spectrum that are related to the physically relevant scenario of dark matter with a finite temperature.
\cite{Konradetal2020} have shown that the Zel'dovich power spectrum for large $k$ can be expanded in a series with terms of order $k^{-3-2m}$, when the initial power spectrum has a cut-off and all moments of the initial velocity potential power spectrum exist.
The first two coefficients in the series \eqref{eq:6} are given by
\begin{equation}
\begin{aligned}
    \mathcal{P}^{(0)}(t) &= 3 (4\pi)^{3/2} \left( \frac{5}{2 \tau_2^2} \right)^{3/2} \mathrm{e}^{- \frac{5}{2 \tau_2^2}}\;,
    \\
    \mathcal{P}^{(1)}(t) &= \frac{(4 \pi)^{3/2}}{28} \frac{\sigma_3^2}{\sigma_2^2} \left( \frac{5}{2 \tau_2^2}\right)^{5/2} \mathrm{e}^{- \frac{5}{2 \tau_2^2}} \left[123 -132 \left( \frac{5}{2 \tau_2^2} \right) + 20 \left( \frac{5}{2 \tau_2^2} \right)^2 \right]\;,
\end{aligned}
\label{eq:37}
\end{equation}
where $\mathcal{P}(k,t) \sim \mathcal{P}^{(0)}(t)k^{-3} + \mathcal{P}^{(1)}(t)k^{-5}$ as $k\to \infty$.
While the first term describes the asymptotic $k^{-3}$ tail very well, the intersection with the second term, of order $k^{-5}$, engenders a time-dependent scale $k_\text{asymp} = \sqrt{\left|\frac{\mathcal{P}^{(0)}}{\mathcal{P}^{(1)}}\right|}$, below which the asymptotics fail to reproduce the Zel'dovich power spectrum. Thus, an analytical description of the latter between this scale and the linear regime was missing.

\begin{figure}
    \centering
    \includegraphics[width = 0.95\textwidth]{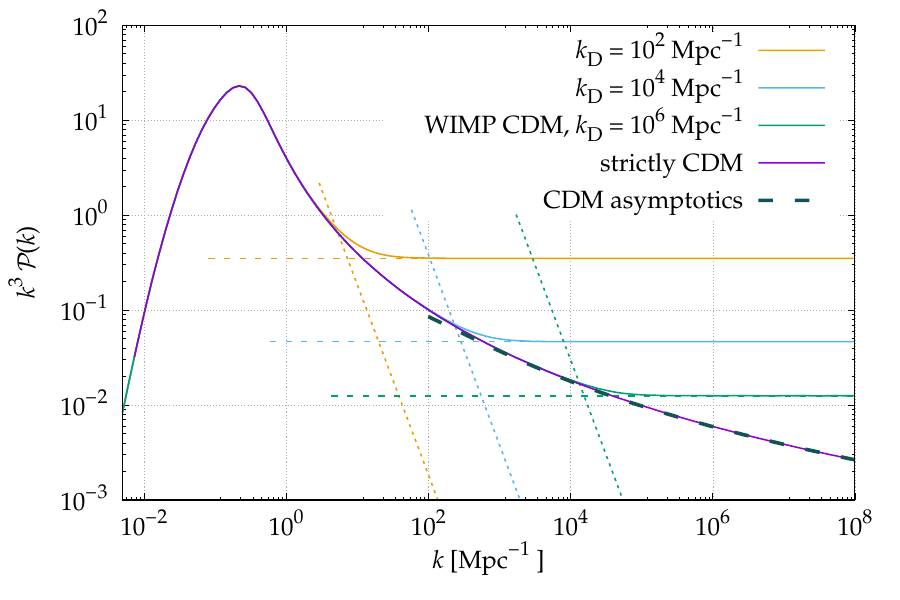}
    \caption{The cold dark matter power spectrum in the Zel'dovich approximation without small scale smoothing (purple line), compared with power spectra that are initially smoothed on small scales. For larger smoothing wave numbers, i.e. smaller smoothing scales, the initially smoothed power spectra align with the unsmoothed spectrum up to higher wave numbers $k$. This implies that the intermediate regime of initially smoothed Zel'dovich power spectra is well approximated by the asymptotics of the exact cold dark matter power spectrum. The Dicus transfer function \citep{Weinberg2008} and the parameters in Table \ref{tab:parameters} are used.}
    \label{fig:dimension-less z=0 comparison to smoothed}
\end{figure}

In Fig. \ref{fig:dimension-less z=0 comparison to smoothed} we show the dimension-less Zel'dovich power spectra of strictly cold dark matter without an initial ultraviolet cut-off, i.e. when dark matter is treated as a non-relativistic Vlasov fluid at all scales (purple line), at redshift $z=0$, together with those of three types of warm dark matter, where the initial (linear) power spectrum is exponentially cut off at a scale $k_D = \{10^2, 10^4, 10^6\}$ Mpc$^{-1}$ by a multiplicative Gaussian $\exp(-k^2/k_D^2)$ (yellow, blue and green line, respectively). The latter corresponds to a power spectrum of WIMP dark matter with particle mass of $\sim100$ GeV \citep{Greenetal2004}. For small and intermediate wave numbers, the Zel'dovich power spectra align with the spectrum of the strictly cold dark matter. At large wave numbers, the spectra become flat, reaching the $k^{-3}$ asymptotics (coloured dashed lines). The wave number where the spectra with finite initial temperature deviate from the strictly cold dark matter spectrum is in all shown cases indicated very well by the intersection of the $k^{-3}$ (coloured dashed lines) and the $k^{-5}$ (dotted dashed lines) asymptotics. In case of the WIMP dark matter, the wave number regime just below validity of the $k^{-3}$ asymptotics, i.e. below $10^4$ Mpc$^{-1}$, is now very well described by the asymptotics of the strictly cold dark matter (thick dashed dark blue line) that we derived in this paper up to order $M=1000$. This is remarkable, since the latter contains information about the asymptotic tail of the initial power spectrum, that is actually cut-off for WIMP dark matter, but ultimately originates from the fact that $\sigma_3^2/k_s^4 \gg \abs{\mathcal{M}[P_\delta^{(i)};5]}$, where the former is computed with a cut-off, and the latter -- without one.

It may be surprising, \emph{prima facie} that the asymptotic tails of the Zel'dovich power spectra with initial ultraviolet cut-off have a higher amplitude at redshift $z=0$ the stronger the ultraviolet cut-off is, i.e. the smaller $k_D$. The reason for this is that the time evolution of the asymptotic amplitude $\mathcal{P}^{(0)}$, \eqref{eq:37}, depends on the product $\sigma_2^2 t^2$. This implies that smaller values of $\sigma_2$ lead to a slower evolution of the asymptotic tail. Since at redshift $z = 0$, all of these amplitudes are decreasing in time due to re-expansion, the spectra with larger $k_D$, i.e. larger $\sigma_2^2$ already progressed further in re-expansion. The time evolution of the amplitude is discussed in detail in \cite{Konradetal2020}.

\section{Summary and discussion}
\label{sec:summary}

In this paper we derived an asymptotic expansion for the cold dark matter power spectrum in the Zel'dovich approximation, at large wave numbers $k$, by using the Mellin transform technique; the result, Eq. \eqref{eqn:whole k dependence extracted}, shows that in addition to the expected $k^{-3}$, $k^{-5}$, \emph{etc.} terms, there are terms that go like negative integer powers $k^{1-n_s}$, accompanied by positive powers of $\log k$. These are a \emph{sine qua non} in any practical use of asymptotic expansions of the Zel'dovich power spectrum, since only by going to high order, can a reasonable accuracy be achieved (see Fig. \ref{fig:late time z=0}).

This differs from the case where the power spectrum contains an ultraviolet cut-off, $k_D$ \citep{Konradetal2020}. In this case, there exists a cut-off and time dependent wave number $k_{\text{asymp}}$ above which the Zel'dovich power spectrum aligns with the $k^{-3}$ asymptotics. However if $k_D \gg k_s$, there is an intermediate regime, where $k_s \ll k \ll k_{\text{asymp}}$. There, physically, the cut-off should have no effect, and for these values of $k$, we indeed see in Fig. \ref{fig:dimension-less z=0 comparison to smoothed} that $\mathcal{P}(k,t)$ follows \eqref{eqn:whole k dependence extracted}, rather than the asymptotic expansion of \cite{Konradetal2020,ChenPietroni2020}.
This already means that for such $k$ one should not use the expansion with a cut-off, but rather treat the cut-off as non-existent. The applicability of an expansion without a cut-off, well below the cut-off in asymptotic expansions is known to exist elsewhere in cosmology, e.g. in the distribution of the astrophysical gravitational-wave background \citep{Ginatetal2019}. We therefore have shown that \eqref{eqn:whole k dependence extracted} is valid even for realistic models of dark matter, which do include some ultraviolet cut-off, as long as one considers scales below $k_{\text{asymp}}$. We therefore conclude that for Zel'dovich power spectra of dark matter with an initial ultraviolet cut-off, the $k \to \infty$ asymptotics are correctly described by the result of \cite{Konradetal2020}, \eqref{eq:6}, rather than \eqref{eqn:whole k dependence extracted}. 
This is because the $m>0$ terms in \eqref{eqn:whole k dependence extracted} appear only due to the poles of the Mellin transform of the initial power spectrum when no initial ultraviolet cut-off is imposed. 
For some intermediate regime of finite values of $k$, however, at wave numbers below which the series \eqref{eq:6} fails to reproduce the warm dark matter Zel'dovich power spectra, warm dark matter and stricly cold dark matter spectra align and the warm dark matter spectra \emph{can} be described by the asymptotics of the cold dark matter spectra \eqref{eqn:whole k dependence extracted} derived in this work. We suspect that the mathematical reason for this is that there might exist an expansion for the warm dark matter spectra in the $k\rightarrow \infty$ limit for which the $m>0$ terms do not vanish strictly, but are suppressed by an $O(k^{-\infty}$ function (i.e. one that falls off more steeply than any power law), with the suppression controlled by the cut-off scale $k_D$. Since such terms would be non-perturbative, they would not appear in the asymptotic expansion \eqref{eq:6}. In between those two regimes, i.e. at wave numbers of about $k_{\text{asymp}}$, the warm dark matter power spectrum follows some interpolation of the two asymptotic expressions. It will be subject to future work to investigate the connection between those two expansions and to formally derive the correct interpolation.

While one has to include many terms in the series \eqref{eqn:whole k dependence extracted} for it to be accurate, knowing the functional form of the $k$-dependence of the power spectrum is of theoretical significance in itself. Evaluating the Zel'dovich power spectrum can be achieved by solving the integral \eqref{eq:KFT zeldovich PS} numerically.
This expansion will help to find out more about the mathematical structure of the Zel'dovich power spectrum in general; we anticipate that once this series is known, in the future one would be able to re-sum it to arrive at more accurate results without needing to go to very high orders. The Zel'dovich power spectrum resembles the zeroth order power spectrum of KFT and is a fundamental building-block of perturbative terms; the result of this paper will therefore guide future work on KFT perturbation theory and thereby on analytical expressions for the non-linear power spectrum of dark matter, and put them on mathematically solid ground. 
Here, we have approximated particle trajectories by straight lines. Due to this Zel'dovich approximation, virialization is not expected to occur. However, virialization is an emergent effect in KFT and thus need not be introduced explicitly. Incorporating gravitational interactions beyond the Zel'dovich approximation will lead to more persistent structures on small scales, since gravity prevents the structures formed in the Zel'dovich approximation from re-expanding. Nonetheless, we have convincing reasons to believe that the shape of the asymptotics will not be changed even by interactions: Taking them into account in a mean-field approximation, we could show that the multiplicative interaction term becomes scale-independent for large $k$ \citep{Bartelmannetal2021, Konradetal2022}. Then, the amplitude of the power spectrum will increase, but the shape will remain unchanged. The analytical result based on the mean-field approximation recovers fully numerical results for non-linear power spectra to an accuracy of a few percent up to wave numbers $k\lesssim 10\,h\,\mathrm{Mpc}^{-1}$. 
In future work we will extend the present calculation beyond the Zel'dovich approximation, and endeavour to find an asymptotic series of the form \eqref{eqn:whole k dependence extracted} for the full cold dark-matter power spectrum.% \citep{Ginatetal2022}.

We did not consider baryonic effects either here, but dark matter only as a back-bone for cosmic structure formation. This is of particular cosmological interest because the question as to which small-scale structures in (warm and cold) dark-matter only universes form is not completely solved yet, e.g.\ in the notorious ``core-cusp'' problem \citep{Geninaetal2018}. While baryonic effects are certain to change the small-scale structures built from dark matter, it is yet unclear which small-scale structures actually arise in dark-matter only universes, as numerical simulations are limited in resolution. We are convinced that this is an important question to solve, prior to incorporating any baryonic effects.

Moreover, as is visible from Fig. \ref{fig:sigmas_for_ns}, the coefficients $\mathcal{P}_m(t,\ln k)$ (or $\mathcal{P}_{mn}(t)$) are extremely sensitive to the spectral index $n_s$. 
We believe that the primordial $n_s$ may leave an imprint on cosmic structure formation at early times. Since the cooling time of the baryonic matter is finite, the formation of dense baryonic structures is expected to set in later than small-scale structures form by stream-crossing in cold dark matter. These small-scale, cold dark-matter structures will in turn influence the mass and time scales of baryonic structure formation. These different time scales might themselves be reflected in observables related to reionization or the recombination epoch. It is thus quite possible that the primordial $n_s$ may be related to observables of cosmic structure formation. It will be the subject of future work to investigate these time scales and predict observables, with initial conditions for our theory set during the matter-dominated era.
In future surveys, if they could be inferred experimentally, they could furnish an additional probe of this cosmological parameter.

\section*{Acknowledgements}
S.K. and M.B. thank Manfred Salmhofer for numerous helpful discussions, and Y.B.G. thanks Vincent Desjacques, Hagai Perets and Mor Rozner for useful comments. S.K. thanks Robert Scheichl for helpful conversations. Y. B. G. is grateful for the generous hospitality of the Institute of Theoretical Physics at Heidelberg University, where some work on this research was done.
This research received funding from the Deutsche Forschungsgemeinschaft (DFG, German Research Foundation) under Germany's Excellence Strategy EXC 2181/1 - 390900948 (the Heidelberg STRUCTURES Excellence Cluster). Y. B. G. is also supported by the Adams Fellowship Programme of the Israeli Academy of Sciences and Humanities.

\section*{Data Availability}
The data underlying this article will be shared on reasonable request to the corresponding authors.

\bibliographystyle{mnras}
\bibliography{main}

\appendix

\section{Mellin transform technique}
\label{appendix:Mellin}
In this section, we briefly sketch the Mellin transform technique as described in \cite{bleistein1975asymptotic}, in-so-far as it is used in this paper. This technique is a powerful tool to derive the asymptotics of functions that are described by integral transforms.
The Mellin transform of a function $f(k)$ is defined as
\begin{equation}
\mathcal{M}\left[f;z \right] \equiv \int_0^{\infty} \mathrm{d}k ~ k^{z-1} f(k)\;.
\end{equation}
The H-transform of a function $f$ with respect to a function $h$ is defined as
\begin{align}
H[f;\lambda]
& \equiv \int_0^\infty \mathrm{d}k ~ f(k) h(\lambda k)
\\
&= \frac{1}{2 \pi \mathrm{i}} \int_{c - \mathrm{i}\infty}^{c+\mathrm{i}\infty} \lambda^{-z} \mathcal{M}[h;z] \mathcal{M}[f;1-z] \mathrm{d}z\;,
\end{align}
where $c \in \mathbb{R}$ lies in the strip of analyticity of $\mathcal{M}[h;z]$ and $\mathcal{M}[f;1-z]$.
Upon defining
\begin{equation}
G(z) \equiv \mathcal{M}[h;z] \mathcal{M}[f;1-z]\;,
\end{equation}
and using the residue theorem, the $H$-transform has the asymptotic expansion as $\lambda \rightarrow \infty$
\begin{equation}
H[f;\lambda]
=
- \sum_{c < \mathrm{Re}~ z < R} \mathrm{Res}\left\{ \lambda^{-z} G(z)\right\} + \frac{1}{2\pi \mathrm{i}}
\int_{R-\mathrm{i}\infty}^{R+\mathrm{i}\infty} G(z) \mathrm{d}z\;,
\end{equation}
where $R$ is chosen such that $\lambda^{-z}G(z)$ has no poles for $\mathrm{Re}~z=R$. The latter integral is $\mathcal{O}(\lambda^{-R})$ as $\lambda \rightarrow \infty$, where $G$ is analytically continued as much as possible, preferably to a meromorphic function of $z$.

To calculate the asymptotics as $\lambda \rightarrow \infty$ explicitly, we assume that, as $t \rightarrow \infty$,
\begin{equation}
h(t) \sim \mathrm{e}^{-dt^{\nu}} \sum_{m=0}^{\infty} t^{-r_m} \sum_{n=0}^{N(m)} c_{mn} \ln^n t\;,
\end{equation}
where $d\ge 0$, $\nu>0$, $\mathrm{Re}~r_m \nearrow \infty$, $0 \le N(m)$ is finite.
We also assume that as $t \rightarrow 0$,
\begin{equation}
f(t) \sim \mathrm{e}^{-qt^{-\mu}} \sum_{m=0}^{\infty} t^{a_m} \sum_{n=0}^{\bar{N}(m)} p_{mn} \ln^n t\;,
\end{equation}
where $q\ge 0$, $\mu>0$, $\mathrm{Re}~a_m \nearrow \infty$, and $0 \le \bar{N}(m)$ is finite. The relevant theorem, quoted here from \cite{bleistein1975asymptotic} for completeness, considers four cases:

\begin{itemize}
\item Case I: $d \neq 0 \neq q$:
\begin{equation}
H[f;\lambda] = o(\lambda^{-R}) \quad \forall R>0\;.
\end{equation}
\item Case II: $d \neq 0$, $q = 0$:
\begin{equation}
H[f;\lambda] \sim \sum_{m=0}^{\infty} \lambda^{-1-a_m} \sum_{n=0}^{\bar{N}(m)} p_{mn} \sum_{j=0}^n \binom{n}{j} (-\ln \lambda)^j \left.\mathcal{M}^{(n-j)}[h;z]\right|_{z=1+a_m}\;.
\end{equation}
\item Case III: $d = 0$, $q \neq 0$:
\begin{equation}
H[f;\lambda] \sim \sum_{m=0}^{\infty} \lambda^{-r_m} \sum_{n=0}^{N(m)} c_{mn} \sum_{j=0}^n \binom{n}{j} \ln^j \lambda \left.\mathcal{M}^{(n-j)}[f;z]\right|_{z=1-r_m}\;.
\end{equation}
\item Case IV: $d = 0$, $q = 0$:
\begin{itemize}
\item $r_m \neq a_n+1$ for all pairs $n,m$, then the resulting asymptotics is simply given by the sum of Cases II and III.
\item When $r_m = a_n+1$ for one or several pairs $n,m$, then one has to consider the residue
\begin{equation}
\mathrm{Res}_{r_m} \left\{ \lambda^{-z} \mathcal{M}[h;z] \mathcal{M}[f;1-z] \right\}\;,
\end{equation}
which follow from the equation below for the residue. Note that the multiplicity (order) of the pole is increased, which produces logarithmic terms even if there weren't any in $f$ or $h$ to begin with.
\end{itemize}
\end{itemize}
In all the above cases, $\mathcal{M}$ denotes the Mellin transform, or its analytical continuation. $\mathcal{M}^{(n-j)}$ denotes the $(n-j)$-th derivative of the Mellin transform, or its analytical continuation.

\section{Asymptotics of the initial momentum correlations}
\label{appendix:asymptotics}

\subsection{Small Scale Asymptotics}
\label{appendix:asymptotics small scale}
We assume that the initial power spectrum $P_{\delta}^{(i)}(k)$ has an asymptotic expansion as $k \rightarrow \infty$ that is given by
\begin{align}
P_{\delta}^{(i)}(k)
&= A \left(\frac{k}{k_s}\right)^{n_s} T^2_D \left(\frac{k}{k_s}\right)
\sim \left(\frac{k}{k_s}\right)^{n_s-4} \sum_{m=0}^{\infty} \left(\frac{k}{k_s}\right)^{-m} \sum_{n=0}^{2} c_{mn} \ln^n \left(\frac{k}{k_s}\right)\;.
\label{eq:a:A:1:1}
\end{align}
Note that with $\kappa = k/k_s$ and $k_s$ as defined in \eqref{eq:12} the argument is actually dimension-less.

The functions $a_1$ and $a_2$ that characterise the initial momentum correlations are given by
\begin{align}
a_1(q) &:= \frac{\xi'_{\psi}(q)}{q} = - \frac{1}{2\pi^2} \int_0^{\infty} \mathrm{d}k \ P^{\text{(i)}}_{\delta} (k) \frac{j_1(kq)}{kq}\;,
\label{eq:a1a2:11:2}
\\
a_2(q) &:= \xi''(q) -  \frac{\xi'_{\psi}(q)}{q} = \frac{1}{2\pi^2} \int_0^{\infty} \mathrm{d}k \ P^{\text{(i)}}_{\delta} (k) j_2(kq)\;.
\label{eq:a1a2:12:2}
\end{align}
let us derive the asymptotics of these functions for small scales, i.e. when $q \rightarrow 0$, by applying the Mellin transform technique.

Let's commence with the function $a_1$ for the large parameter $x^{-1} \equiv k_s^{-1}q^{-1}$. We start by transforming the integral expression for $a_1$ by using $\kappa = k/k_s$
\begin{equation}
a_1(q)
= - \frac{1}{2\pi^2} \int_0^{\infty}\mathrm{d}k ~  P_{\delta}^{(i)}(k) \frac{j_1(\kappa x)}{\kappa x}
= - \frac{1}{2\pi^2} x^{-1} k_s \int_0^{\infty}\mathrm{d}\kappa ~  P_{\delta}^{(i)}(k_s \kappa x^{-1}) \frac{j_1(\kappa)}{\kappa}
= - \frac{1}{2\pi^2} x^{-1} k_s \int_0^{\infty}\mathrm{d}\kappa ~  A \left(\frac{\kappa}{x}\right)^{n_s} T^2_D\left( \frac{\kappa}{x}\right) \frac{j_1(\kappa)}{\kappa}
\end{equation}
Since the spherical Bessel function $j_1$ has a convergent Taylor series at $\kappa=0$,
\begin{equation}
    \frac{j_1(\kappa)}{\kappa} = \sum_{m=0}^{\infty} \kappa^{2m} \frac{(-1)^m}{(2m+3)(2m+1)!},
\end{equation}
and $T^2_D$ has a power law tail with logarithms \eqref{eq:a:A:1:1}, the resulting asymptotics is given by Case IV of the appendix \ref{appendix:Mellin}. Since we assume $0 < n_s < 1$, the exponents of the series of $P_{\delta}^{(i)}$ in \eqref{eq:a:A:1:1} are not integers, which implies that the asymptotic series is a sum of Cases II and III:
\begin{equation}
a_1(q)
\sim - \frac{1}{2\pi^2} \sum_{m=0}^{\infty} x^{2m} \frac{(-1)^m}{(2m+3) (2m+1)!} \mathcal{M}\left[P_{\delta}^{(i)};1+2m\right]
- \frac{1}{2\pi^2} x^{3-n_s} \sum_{m=0}^{\infty} x^{m} \sum_{n=0}^{2} \frac{c_{mn}}{A} \sum_{j=0}^n \binom{n}{j}  (- \ln x)^j \left.\mathcal{M}^{(n-j)} \left[j_1;z\right] \right|_{z=n_s-4}\;,
\end{equation}
where we have defined
\begin{equation}
    \mathcal{M}\left[P_{\delta}^{(i)};z\right] \equiv A k_s \int_0^{\infty} \kappa^{n_s+z-1} T^2_D(\kappa)\;\mathrm{d}\kappa\;.
\end{equation}
Note that for $z=2m-1$, these terms are scaled versions of the analytical continuation of the moments of the initial velocity power spectrum \eqref{eq:4},
\begin{equation}
    \mathcal{M}\left[P_{\delta}^{(i)};2m-1\right] = 2 \pi^2 \frac{\sigma_m^2}{k_s^{2m-2}}\;.
\end{equation}
Similarly, we get for $a_2$ as $q \rightarrow 0$
\begin{align}
a_2(q) &= \frac{1}{2\pi^2} \int_0^{\infty}\mathrm{d}k ~  P_{\delta}^{(i)}(k) j_2(kq)
=
\frac{1}{2\pi^2} x^{-1} k_s \int_0^{\infty}\mathrm{d}\kappa ~  P_{\delta}^{(i)}(k_s \kappa x^{-1}) j_2(\kappa)
\\
& \sim \frac{1}{2\pi^2} \sum_{m=0}^{\infty} x^{2m+2} \frac{(-1)^m (2m+2)}{(2m+5)(2m+3)!} \mathcal{M}\left[P_{\delta}^{(i)};3+2m\right]
+
\frac{1}{2\pi^2} x^{3-n_s} \sum_{m=0}^{\infty} x^{m} \sum_{n=0}^{2} \frac{c_{mn}}{A} \sum_{j=0}^n \binom{n}{j}  (- \ln q)^j \left.\mathcal{M}^{(n-j)} \left[j_2;z\right] \right|_{z=n_s-3}\;.
\end{align}

\subsection{Large Scale Asymptotics}
\label{appendix:asymptotics large scale}
We assume that the initial power spectrum $P_{\delta}^{\text{(i)}}$ has an asymptotic expansion as $k \rightarrow 0$ that is given by
\begin{equation}
    P_{\delta}^{\text{(i)}} \sim \kappa^{n_s} \sum_{m=0}^{\infty} p_m \kappa^m\;,
\end{equation}
with $\kappa = k/k_s$ and coefficients $p_m$ that have units of volume. We consider again the integral expressions of $a_1$ and $a_2$ in terms of the dimensional integration variable $\kappa$ and $x = k_s q$,
\begin{align}
    a_1(q) &= -\frac{1}{2\pi^2} k_s \int_0^{\infty} \mathrm{d}\kappa \, P_{\delta}^{\text{(i)}}(k_s \kappa) \frac{j_1(\kappa x)}{\kappa x}\;,
    \\
    a_2(q) &= \frac{1}{2\pi^2} k_s \int_0^{\infty} \mathrm{d}\kappa \, P_{\delta}^{\text{(i)}}(k_s \kappa) j_2(\kappa x)\;.
\end{align}
Since the spherical Bessel functions $j_n(x)$ are oscillating for $x \rightarrow \infty$, in order to compute the asymptotics of $a_1$ and $a_2$ for $q \rightarrow \infty$, Case II of the Mellin transform technique applies for both integrals. Thus, as $x = k_s q \rightarrow \infty$,
\begin{align}
    a_1(q) &\sim -\frac{1}{2\pi^2} \sum_{m=0}^{\infty} x^{-1-n_s-m} k_s p_m \mathcal{M}\left[j_1;n_s+m\right]
    = - \frac{x^{-1-n_s}}{2\pi^2} k_s p_0 \sqrt{\pi} \frac{2^{n_s-2} \Gamma \left( \frac{n_s+1}{2}\right)}{\Gamma\left( \frac{4-n_s}{2} \right)} + \mathcal{O}\left(q^{-2-n_s}\right)
    \;,
    \\
    a_2(q) &\sim \frac{1}{2\pi^2}\sum_{m=0}^{\infty} x^{-1-n_s-m} k_s p_m \mathcal{M}\left[j_2;1+n_s+m\right]
    = \frac{x^{-1-n_s}}{2\pi^2} k_s p_0 \sqrt{\pi} \frac{2^{n_s-1} \Gamma \left( \frac{n_s+3}{2}\right)}{\Gamma\left( \frac{4-n_s}{2} \right)} + \mathcal{O}\left(q^{-2-n_s}\right)\;.
\end{align}

\subsection{Computing the analytic continuation of the Mellin transform}
In order to actually evaluate the coefficients appearing in the small scale asymptotics of $a_1$ and $a_2$, we need to compute the analytical continuation of the moments of the initial power spectrum $\mathcal{M}\left[P_{\delta}^{(i)};2m-1\right]$ into the right-hand half-plane numerically.
We assumed that $P_{\delta}^{(i)}$ has an asymptotic expansion as $k \rightarrow \infty$ that is given by
\begin{equation}
P_{\delta}^{(i)} \sim \sum_{m=0}^{\infty} \kappa^{-r_m} \sum_{n=0}^{2} c_{mn} \ln^n \kappa\;.
\end{equation}
The analytic continuation of its Mellin transform is calculated by first defining the function
\begin{equation}
h_p(\kappa) := \begin{cases}
\sum_{\mathrm{Re}~ r_m < p} \kappa^{-r_m} \sum_{n=0}^{N(m)} c_{mn} \ln^n \kappa\;, &\quad \text{for } \kappa \ge 1\;,
\\
0\;, &\quad \text{else}\;.
\end{cases}
\end{equation}
Then, by defining
\begin{equation}
    \tilde{P}_{\delta}^{(i)}(\kappa) := A \kappa^{n_s} T_D^2\left(\kappa\right) = P_{\delta}^{(i)}(k_s \kappa)\;,
\end{equation}
we get for the following expression
\begin{align}
\mathcal{M}\left[P_{\delta}^{(i)}, z\right]
&= k_s \int_0^{\infty} \mathrm{d}\kappa ~ \kappa^{z-1} \tilde{P}_{\delta}^{(i)}(\kappa)
=
k_s \int_0^{1} \mathrm{d}\kappa ~ \kappa^{z-1}\tilde{P}_{\delta}^{(i)}(\kappa) + k_s \int_1^{\infty} \mathrm{d}\kappa ~ \kappa^{z-1} \tilde{P}_{\delta}^{(i)}(\kappa)
\\
&=
k_s \int_0^{1} \mathrm{d}\kappa ~ \kappa^{z-1} \tilde{P}_{\delta}^{(i)}(\kappa)
+ k_s \int_1^{\infty} \mathrm{d}\kappa ~ \kappa^{z-1} \left[\tilde{P}_{\delta}^{(i)}(\kappa)-h_p(\kappa) \right]
+ k_s \int_1^{\infty} \mathrm{d}\kappa ~ \kappa^{z-1} h_p(\kappa)\;,
\end{align}
where the first two integrals can be calculated numerically and the last one is given by its analytical continuation
\begin{equation}
\int_1^{\infty} \mathrm{d}\kappa ~ \kappa^{z-1} h_p(\kappa) =
\sum_{\mathrm{Re}~r_m < p} \sum_{n=0}^{N(m)} \frac{(-1)^{n+1} c_{mn} n!}{(z-r_m)^{n+1}}\;.
\end{equation}

\section{Neglecting other orders in the asymptotics}
\label{appendix: neglect k^-5}
\begin{figure}
    \centering
    \includegraphics[width=1.0\textwidth]{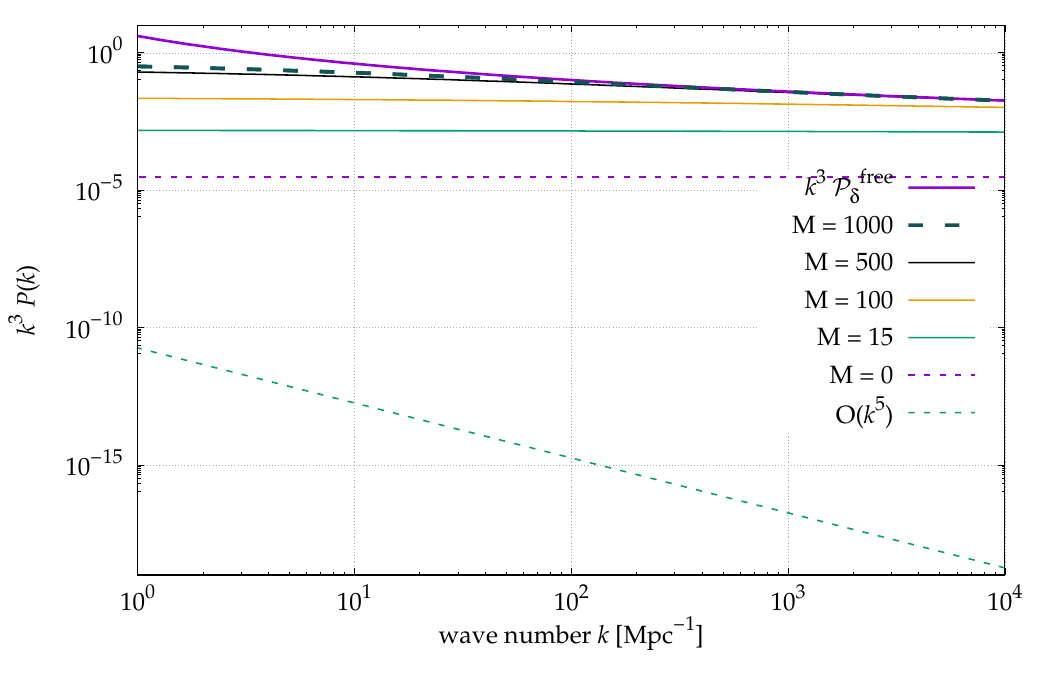}
    \caption{The cold dark matter power spectrum in the Zel'dovich approximation at $z = 0$ (solid purple line), compared with the analytical prediction of the asymptotics \eqref{eqn:late time Pm}, up to high orders. On the scales shown here, the $k^{-3}$ asymptotics (dashed purple line) underestimate the dimension-less power spectrum by approximately an order of magnitude and the $\ord{k^{-5}}$ term is completely negligible. The Dicus transfer function and the parameters in table \ref{tab:parameters} are used.}
    \label{fig:appendix dimension-less spectrum}
\end{figure}

While for dark matter with an initial ultraviolet cut-off the $k^{-5}$ asymptotics typically dominate the $k^{-3}$ asymptotics on small scales (see figure \ref{fig:dimension-less z=0 comparison to smoothed}), this is not the case for strictly cold dark matter. To compute the asymptotics for the Zel'dovich power spectrum of strictly cold dark matter, we therefore exclusively added terms from the sum \eqref{eq:23}, i.e. we did not add the order $k^{-5}$ term, although $M=1000$ implies that we went up to order $-M(1-n_s) = -33.3$. Nevertheless, as can be seen in figure \ref{fig:appendix dimension-less spectrum}: at today's redshift $z=0$ this term is negligible at small scales, because the logarithms in the numerator almost cancel the powers of $k$ in the denominator.

\section{Large $m$}
\label{appendix:large m}
The integral in \eqref{eqn:late time Pm} is a double integral over $y$ and $\mu$. Simply evaluating it numerically in \verb+Mathematica+ works only up to $m \lesssim 50$. For larger $m$, one might presume na\"{i}vely that a simple Laplace-type approximation would be appropriate, but this is true only for the $y$ integral.

Let us start, therefore, with the $y$ integral, and write
\begin{equation}
    \mathcal{P}_m = \frac{1}{m!}\int_0^\infty \mathrm{d}y \int_{-1}^1 \mathrm{d}\mu ~y^2 e^{m(1-n_s) \ln y - (2\mu^2 + 1)y^2/30} \left[\frac{\xi}{\sigma^2}\right]^m
\end{equation}
This is a moveable saddle problem, and therefore we need to re-scale $y$ appropriately before proceeding. The exponent
\begin{equation}
    m(3 - n_s) \ln y - \frac{2\mu^2 + 1}{30} y^2
\end{equation}
is stationary when
\begin{equation}
    y^2 = m \frac{15(3-n_s)}{2\mu^2 + 1},
\end{equation}
i.e. the stationary point is $\sim \sqrt{m}$. So, let us change variables to $y = u\sqrt{m}$, and define
\begin{equation}
    u_0(\mu) \equiv \sqrt{\frac{15(3-n_s)}{1+2\mu^2}}.
\end{equation}
$\mathcal{P}_m$ then becomes
\begin{align}
    \mathcal{P}_m & \sim \frac{1}{m!} \int_{-1}^1\mathrm{d}\mu \int_0^\infty \mathrm{d}u~m^{(3+m(3-n_s))/2}u_0^2 \left[\frac{\xi(y = \sqrt{m}u_0(\mu)/k\tau_2)}{\sigma^2}\right]^m \exp\left[\frac{m(3-n_s)}{2} \left( \ln(u_0^2) - 1\right)\right] \exp\left[-m\frac{1+2\mu^2}{15}(u-u_0)^2 + \ldots \right]\\ &
    \sim \frac{m^{\frac{3+m(3-n_s)}{2}}e^{-m(3-n_s)/2}}{m!} \int_{-1}^1\mathrm{d}\mu~u_0^{2+m(3-n_s)} \left[\frac{\xi(y = \sqrt{m}u_0(\mu)/k\tau_2)}{\sigma^2}\right]^m \int_{-\infty}^\infty \mathrm{d}u~ e^{-\frac{m(1+2\mu^2)}{15}(u-u_0)^2} \\ &
    = \frac{m^{1+\frac{m(3-n_s)}{2}}e^{-m(3-n_s)/2}\sqrt{15\pi}}{m!} \int_{-1}^1\mathrm{d}\mu~\frac{u_0^{2+m(3-n_s)}}{\sqrt{1+2\mu^2}} \left[\frac{\xi(y = \sqrt{m}u_0(\mu)/k\tau_2)}{\sigma^2}\right]^m,\label{eqn:y integral done, mu to go}
\end{align}
where, in moving to the penultimate line from the one above it, we used Laplace's method for $m \gg 1$.

Equation \eqref{eqn:y integral done, mu to go} can be used to evaluate $\mathcal{P}_m$ numerically with \verb+Mathematica+ up to $m \lesssim 90$ -- after that, the power of $u_0$ and $\xi$ in the integral becomes too large. Upon changing $\mu$ to $x$, Eq.~\eqref{eqn:y integral done, mu to go} may be written as
\begin{equation}
    \mathcal{P}_m \sim \frac{ m^{1+\frac{m(3-n_s)}{2}}e^{-m(3-n_s)/2}\sqrt{15\pi}}{m!} \int_{-1}^1 \mathrm{d}x ~\frac{u_0^2(x)e^{m \varphi(x)}}{\sqrt{1+2x^2}},
\end{equation}
where
\begin{equation}
    \varphi(x) \equiv (3-n_s)\ln u_0(x) + \ln\left[\frac{\xi(y = \sqrt{m}u_0(x)/k\tau_2,x)}{\sigma^2} \right].
\end{equation}
Laplace's method works only if $\varphi$ has variations of order one about it maximum $x=0$. However, here, $\varphi$ is almost constant, for the parameters in table \ref{tab:parameters} and, e.g. $k \gtrsim 10$. For $80 \lesssim m \lesssim 1000$, $m\varphi$ is still almost constant for $x \in [-1,1]$ (i.e. $m|\varphi''(0)|x^2 \ll 1$), and therefore a Laplace-type approximation would be inaccurate. Instead, we simply approximate $\varphi(x) \approx \varphi(0) + \varphi''(0)x^2/2$, and write
\begin{align}
    \mathcal{P}_m & \approx \frac{^{1+\frac{m(3-n_s)}{2}}e^{-m(3-n_s)/2 + m\varphi(0)}\sqrt{15\pi}}{m!} \int_{-1}^1 \mathrm{d}x ~\frac{u_0^2(x)}{\sqrt{1+2x^2}}\left(1+m\varphi''(0) \frac{x^2}{2}\right) \\ &
    =
    \frac{m^{1+\frac{m(3-n_s)}{2}}e^{-m(3-n_s)/2 + m\varphi(0)}\sqrt{15\pi}}{m!}\left[15(3-n_s)\right]\left[\frac{2}{\sqrt{3}} + \frac{m\varphi''(0)}{2}\left(\frac{\textrm{arcsinh}(\sqrt{2}}{\sqrt{2}} - \frac{1}{\sqrt{3}}\right)\right]\label{eqn:large m late time}
\end{align}

For figure \ref{fig:late time z=0}, we use \eqref{eqn:y integral done, mu to go} for $40<m\leq 85$, and for $m > 85$ we use \eqref{eqn:large m late time}. Finally, let us remark that for $n_s < 1$, Stirling's approximation for $m!$ implies that the series \eqref{eqn:P_z m series} diverges very slowly, for fixed $k$, but is of course still asymptotic as $k \to \infty$.

% Don't change these lines
\bsp	% typesetting comment
\label{lastpage}
\end{document}